# Molecular aggregation in liquid water: Laplace spectra and spectral clustering of H-bonded network

**Imre Bakó**[*,1], **Dániel Csókás**[1], **Szilvia Pothoczki**[2],

[1] Research Centre for Natural Sciences, H-1117 Budapest, Magyar tudósok körútja 2., Hungary

[2] Wigner Research Centre for Physics, H-1121 Budapest, Konkoly Thege M. út 29-33., Hungary.

*E-mail: bako.imre@ttk.mta.hu; Phone: +36 1 382 6981

**ABSTRACT**

Application of the spectral clustering method based on the analyses of the Laplace matrix is an acceptable indicator of the global properties of H-bonded network. The first peak of the Laplace spectra contains six eigenvalues. These results suggest that six communities are always formed in our simulated systems independently of the number of molecules in the cubic box. We showed that the H-bonded environment on the surface of the clusters is different from what can be found inside of the clusters. The fraction of four-coordinated molecules is significantly larger in the case of surface molecules. Our work emphasizes that the periodic boundary conditions always cause clustering in the system.

1. **Introduction**

Graph theory has been used for describing topological properties of networks in physics, chemistry, biology and social studies. [1-12] Large number of interactions can be mapped into graphs containing only elemental building blocks such as vertices and edges. Systems which can form extended network established by H-bonds are one of best candidates for spectral graph analyses. Because the interactions between molecules can be considered to be the edges, while the molecules themselves can be the vertices of a graph map. The graphs derived from H-bond network consists of eigenvalues and eigenvectors of their matrices, such as its adjacency matrix or Laplacian matrix. This network-specific approach to characterize the connectivity in H-bonded systems in depth have been already utilized in earlier studies of water and aqueous solutions. [13-21] In our recent study, on pure water forming a 3D space filling percolated network, a well-defined gap was detected in the spectrum at low eigenvalues, and this gap becomes smaller as temperature increases. [20]

This work focuses on revealing the properties of the Laplace spectra for TIP4P/20005 water model [22] to find connection with the structural features of real liquid by means of utilizing the spectral clustering method [23-32]. This approach is based on the eigenvector decomposition of a Laplacian matrix and treats the data clustering as a graph partitioning problem without making any assumption of the form of the data clusters. The relevant eigenvectors are the ones that correspond to the smallest several eigenvalues of the Laplacian except for the smallest eigenvalue which will have a value of 0. This method works well in those cases when smaller eigenvalues groups together and this group are separated with a gap from larger eigenvalues. It has already been shown that the presence of near-zero eigenvalues generally indicates the existence of strong communities, or nearly disconnected components. [23,30,32] To identify the member of the community, which consisted of the nodes belonging to the cluster with larger internal connectivity than external one, is an important task for complex networks which aids us to elucidate the connection among the important modules of the whole network.

The spectral clustering method is mainly based on the eigenvectors (and eigenvalues), thus indirectly the solidity of the conclusions concerning the structural aspect of liquid water largely depends on their reliability. For that reason, we intend to investigate the following:

1. System size dependence of the Laplace spectra of Hydrogen bond's connectivity matrix.
2. Effects of the coordination number on the Laplace spectra through various high symmetry crystal systems. These are cubic diamond, face-centered cubic (fcc), body-centered cubic (bcc), cubic and hexagonal ice.
3. Effects of the periodic boundary conditions on the Laplace spectra with terminating them one by one.

The main aim is to identify by the help of spectral clustering method such types of molecular aggregations in pure liquid water systems, where the inner cluster bond density larger than the inter cluster bond. Thus, a bridge can be established between spectral graph properties and one of unique properties of water, that is the existence of two different phases in water with significantly different densities. Accordingly, the structure of liquid water could be divided into two different forms of the liquid: a high-density liquid (HDL) with less tetrahedral symmetry and a low-density liquid (LDL) with more tetrahedral symmetry.[33-45] It was suggested that the density difference between the two phases is about 20 %.[33] A significant aspect of this theory is that this liquid-liquid transition occurs between these two liquid phases under supercooled conditions, but the fluctuations are affecting the properties of liquid water under ambient conditions. There are several experimental works (e.g. X-ray absorption, emission spectra, X-ray diffraction, Raman spectroscopy) providing evidence for the existence of these two states in liquid water under ambient conditions. [34-40]

Based on molecules dynamics simulation applying the detection method (local structure index, tetrahedrality, Voronoi Void, ring distribution) for these different density phases was mainly based on some types of local order parameters. [41-45] The ratio of amount of locally tetrahedral ordered phase of the higher entropy phase was estimated to be in the range of 1:2.5-3.0, and this value was found to be dependent on the temperature. [37,39, 44, 45] However, the question remains open, because also many experimental and theoretical studies dispute this statement. [46-52]

It is worth emphasizing in this context that in contrast to these local approaches in our study a global method was utilized by the application of spectral clustering algorithm, which assists in scrutinizing the existence of the above mentioned two liquid phases with different densities. The demonstrated differing density can be regarded as a moderate microheterogeneity in the studied systems. The quintessential question is whether this observed clustering is genuinely an inherent property or perturbation of the studied liquid systems.

2. Methods

**2.1 The Network Spectra**

The structure of a network can be fully characterized with the adjacency or the combinatorial Laplace (L) matrices. The Laplace matrix can be defined as follows:

$$L_{ij} = k_i \delta_{ij} - A_{ij} \qquad (1)$$

where $k_i$ is the number of (hydrogen) bonded neighbors of the 'i'-molecule; $\delta_{ij}$ is the Kronecker delta function and $A_{ij}$=1 if a bond exists between the node i and j.

It is known that the Laplacian matrix is positive semidefinite and has nonnegative eigenvalues. Furthermore, 0 is always an eigenvalue of L and the multiplicity of the eigenvalue 0 is equal to the number of the connected components of the graph. It is also known that the Laplace spectra of a system is the union of its connected components (monomer, dimer…largest cluster).

Several authors have studied the relationship between the eigenvector corresponding to the second smallest eigenvalue (λ2) and the graph structure; detailed reviews can be found in the literature. [53-57] The most important theorem connected to the second smallest positive eigenvalue (Fiedler eigenvalue) of a Laplacian is known as the Cheeger inequality [58-60]

$$\frac{\lambda_2}{2} < h(G) > \sqrt{2\lambda_2} \qquad (2)$$

where h(G) is the Cheeger constant of a graph G.

Another prominent feature for the eigenvalue problem of the L matrix is revealed by the structure and localization of components of the eigenvectors. We can characterize these eigenvectors ($\lambda_j$) by using the inverse participation ratio (I), as defined by the following equation:

$$I_j = \sum_i V_{ij}^4 \qquad (3)$$

where $V_{ji}$ is the i-th element of the j-th eigenvector. It has already been also shown that 'I' ranges from the minimum value of 1/N, corresponding to the eigenvector distributed equally on all nodes, to a maximum of 1 for a vector with only one nonzero component.

## 2.2 Simulation Details

All of the molecular dynamics simulations using rigid non-polarizable water models TIP4P/2005 [22] with different system site N=256, 1000, 4000, 8000 have been performed by the GROMACS simulation package [61] (version 5.1.1). The leap-frog algorithm has been used for integrating Newton's equations of motion, with a time step dt=1 fs. The box lengths correspond to the experimental density of water at 298 K.

Diamond's cubic structure is in the Fd3m space group, and the unit cell was composed of 8 atoms. All of the atoms have 4 nearest tetrahedral coordinated carbon neighbors. Each side of the primitive unit cell is 3.57 Å. The Laplace spectra of different size of cells (nxnxn with n=2,3,4,5,6,7; and nxmxm with n=9 and m= 7) was investigated. For calculating the Laplace spectra of the connectivity matrix an in-house code was used. Additionally, two other well-known cubic crystals, namely the face centered cubic (fcc) and the body centered cubic (bcc) with nxnxn (n = 8 and 12) unit cell were studied. The coordination numbers are 8 in fcc and 12 in bcc crystals, respectively. We calculated also the Laplace spectra of the H-bond network of cubic ($I_c$) and hexagonal ice ($I_h$). In these ice structures the oxygen of water molecules is tetrahedral coordinated. For hexagonal ice the unit cell is *orthorhombic* cell with lattice constants *a*=8.9845 Å, *b*=7.7808 Å, and *c*=7.3358 Å. The cubic ice has a cubic unit cell with lattice parameter a=6.35 Å. Periodic boundary conditions (pbc) were applied three different ways: in xyz, in xy, and only in x direction.

In the present study, we applied the following types of H-bonds definitions to ensure that our results are independent of the definition:
  (1) r(O···H) < 2.5 Å and Eij< -12.0 kJ/mol
  (2) r(O···H) < 2.5 Å and O…OH angle is smaller than 30 degree
  (3) r(O···O) < 3.5 Å and Eij< -12.0 kJ/mol
  (4) r(O···O) < 2.5 Å and O…OH angle is smaller than 30 degree.

Two different types of order parameters were applied. One of them is the tetrahedral order parameter, which was originally proposed by Chau et al. [62.] and rescaled by Errington et al. [63]. This quantity is defined for i-th oxygen atoms as

$$q_i = 1 - \frac{3}{8}\sum_{j}^{3}\sum_{k=j+1}^{4}\left(\cos(\theta_{jik}) + \frac{1}{3}\right)^2 \quad (4)$$

where $\theta_{jik}$ is the angle formed by the lines connecting to the central $O_i$ atom to its closest neighbours $O_j$ and $O_k$. The local structure index was calculated using the following formula

$$LSI = \frac{1}{N}\sum_{j}^{N}(\Delta(j) - \Delta_{ave})^2 \quad (5)$$

where

$$\Delta(j) = r(j+1)_{OO} - r(j)_{OO} \quad (6)$$

in this Equation the $r(j)_{OO}$ is the j-th closest O distance from the central "i" oxygen. In the original LSI definition $r(j)max < 3.7$ Å). This quantity is related to the extent of the gap between the first and second hydration shell in liquid water. We additionally calculated a modified version of LSI($n_{hb}$) quantity in which we took into account only the H-bonded neighbours of i-th water. All of the above-mentioned quantities were calculated on independent 500 configurations separated with 100 ps from each other.

## 3. Results and discussion

### 3.1 The Laplace spectra of liquid water

In our earlier studies we showed that for liquid water a well-defined first peak exists at low eigenvalues of the Laplace spectra, which shifted to lower λ values as the temperature increases.[20] The position of this first positive eigenvalue (according to the Cheeger inequalities, see eq.2.) can be considered as metrics for describing the "distance" (that is, number of bonds that need to be broken) from the percolation transition. This first positive eigenvalue always belongs to the first peak. After this peak at least one additional peak separated from the first one can be observed. The Laplace spectra applying different H-bond definitions at lower λ region (Fig. 1c) supports that the pattern is independent of the H-bond definition. For this only results obtained from def. (1) are presented in the following sections.

In order to clarify how the size of the simulation box affects our previous results the Laplace spectra of TIP4P/2005 water model was calculated as a function of the system size. These functions are presented in Fig. 1a and are enlarged at the small λ eigenvalues in Fig. 1b. Independently of the system size a well-defined first peak exists. After this peak a notable gap can be detected. Considering the larger simulation boxes, for the system of 4000 molecules two sharp peaks, while for the system of 8000 molecules three sharp peaks emerge after the gap.

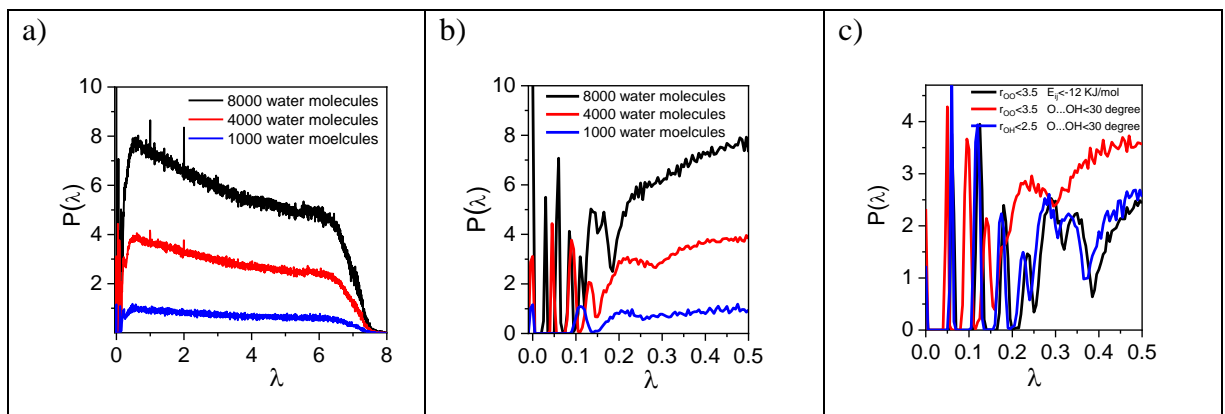

**Figure 1.** Laplace spectra of TIP4P/2005 water molecules as a function of system size a) at the entire λ region, b) only at small λ values and c) only at small λ values applying different H-bond definitions for 4000 water molecules.

The eigenvalues of the Laplace spectra of selected configurations were presented for systems of 1000 (Fig. 2a) and 8000 water molecules (Fig. 2b). In the case of the smaller systems (1000 molecules), the first six eigenvalues form a distinct group. For the larger system (8000 molecules) more groups appear and arrange a kind of plateau-like way. The first six-member group is followed by a 12- and an 8-membered groups. The number of the groups which are composed of the eigenvalues increases with the system size. It is worth mentioning that for system of 4000 molecules two plateau levels were derived containing 6 and 12 eigenvalues (not shown). An additional proof for existence of 6 eigenvalues belonging to the first peaks are presented in Fig. 2c. This figure shows that the difference between successive eigenvalues in the range of 2 to 7 is at least an order of magnitude smaller than the difference between the 8th and seventh eigenvalues. These differences are smaller than the second (Fiedler) eigenvalue. It can be established that the first group of eigenvalues correspond to the first peak of Laplace spectra, the second group to the second peak and so on (c.f. Fig. 1b).

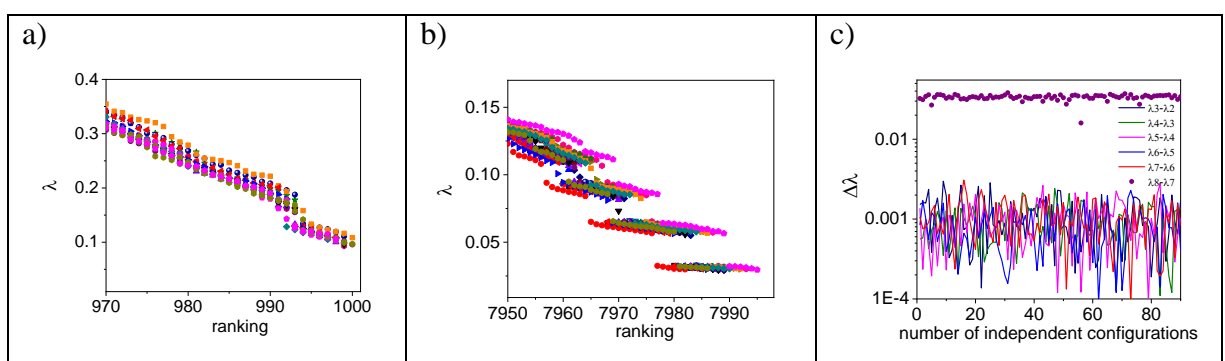

**Figure 2.** The eigenvalues of the Laplace spectra a) for system of 1000 water molecules and b) for system of 8000 water molecules for selected configuration for selected configuration. The different colour is corresponding to the different configuration. c) The $\lambda_i - \lambda_{i-1}$ (i=<8) difference of the Laplacian for the investigated systems.

The second eigenvalue ($\lambda_2$) or in other words the first positive eigenvalue also depends on the system size: 0.191, 0.101, 0.046, 0.036 for system of 256, 1000, 4000 and 8000 water molecules, respectively. That is, $\lambda_2$ demonstrates an inverse proportional change with respect to the number of the molecules in the simulation box.

Inverse participation ratio (IRP) of the eigenvectors was calculated according to Eq.3. The results belonging to the larger eigenvalues in Fig. 3a, while the smaller eigenvalues in Fig. 3b are presented. The IRP of the eigenvectors corresponding to the first and second peaks (ranking range ~7975-7995) are about 2-3 times as large as the minimum value (1/N). This means that these eigenvectors carry information about a large number of molecules. There are several eigenvectors at $\lambda=0$, where the IRP is very large (upper right corner of Fig. 3b). Numerically, IRP in this region is larger than 1/N. These eigenvectors are localized, and only represent a small number of molecules (small clusters, monomer, dimer, etc.) in the system. Union of these connected clusters are the total Laplace spectra of system.

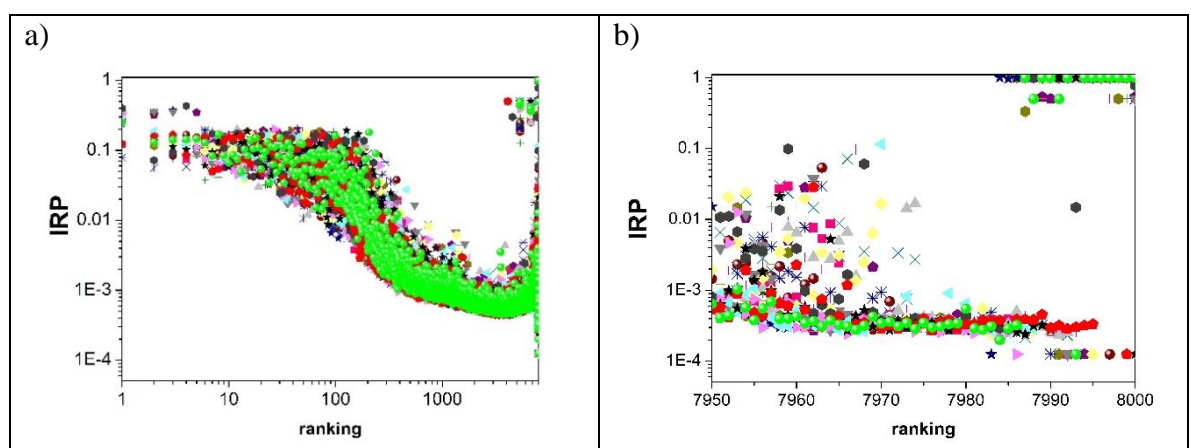

**Figure 3.** Inverse participation ratio (IRP) of the eigenvectors a) for system of 1000 water molecules and b) for system of 8000 water molecules.

In short, independently of the system size the first peak of the Laplace spectra contains six eigenvalues. This result clearly demonstrates that these properties are not related to the finite size effect. One of the possible factors to which this common feature can be attributed is the coordination number. It is, therefore, crucial to carefully examine other systems with higher symmetry properties than the four-coordinated water molecules due to hydrogen bonding.

### 3.2 The Laplace spectra of crystal systems

The Laplace spectra was calculated for some highly symmetric crystal structures in order to shed light on the inherent nature of its band-like structure. Furthermore, the system size dependence was also monitored in all cases. Firstly, cubic diamond was one of the chosen subjects as homogenous, isotropic crystal because similar to water, and having four tetrahedral coordinated neighbors. The resultant spectra are presented in Figure 4a focusing on the low $\lambda$ eigenvalues region up to 2.5. The main feature of the pattern appears the same, but slightly shifted depending on the system size. The multiplicity of the first ($\lambda_2$) and the second nonzero eigenvalue are 6 and 12, respectively. The system size dependence is also noticeable considering the behavior of the $\lambda_2$ and $\lambda_8$-$\lambda_7$, the latter corresponds to the magnitude of the gap between the first and second group of eigenvalues. Fig. 4b shows the results as a function of system size (nxnxn). Both quantity decreases as the system size increases. These results are in accordance with those previously presented for liquid water.

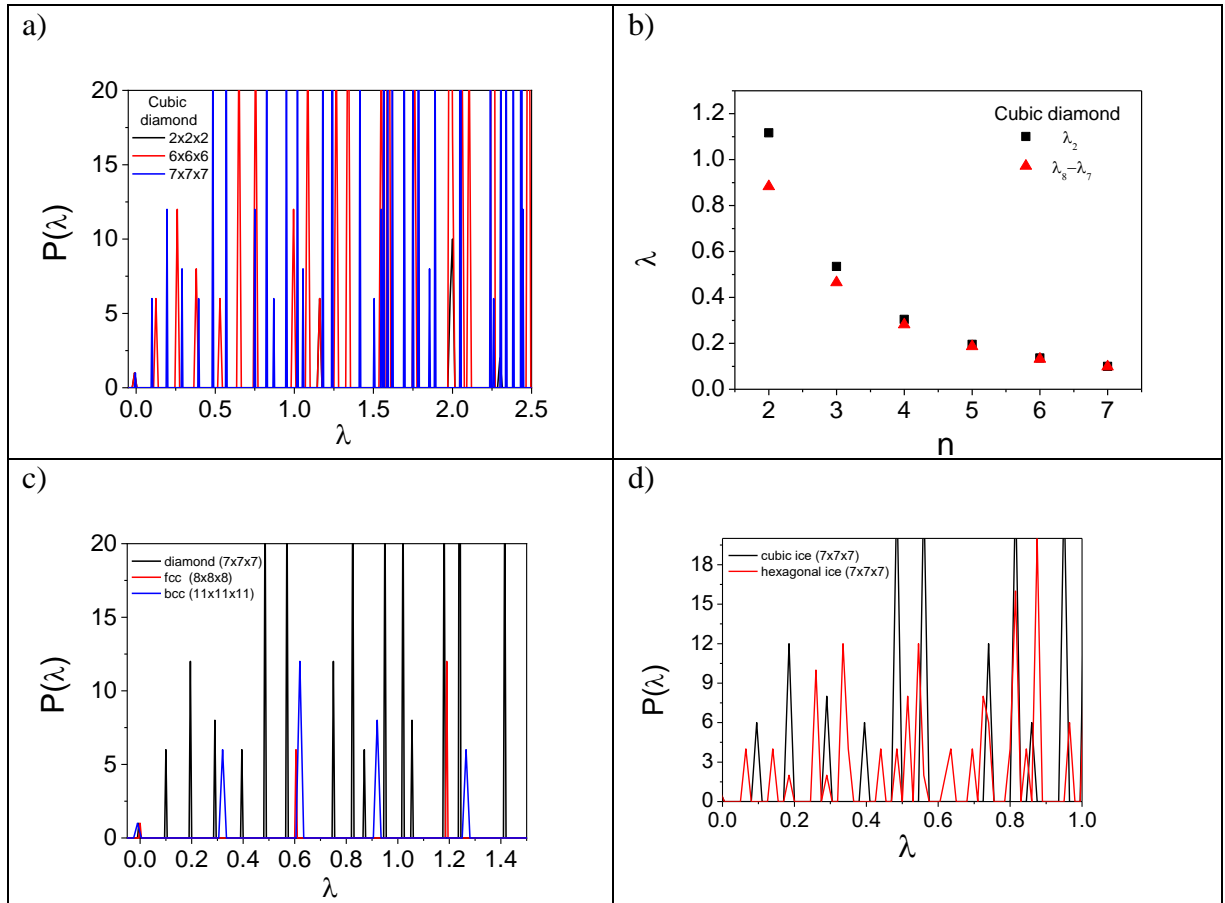

**Figure 4.** a) The Laplace spectra of cubic diamond system with different system sizes. b) The $\lambda_2$ and $\lambda_7$-$\lambda_6$ of cubic diamond systems as a function of the system size. c) The Laplace spectra of fcc and bcc systems. d) The Laplace spectra of cubic ($I_c$) and hexagonal ice ($I_h$) systems.

As the next step the Laplace spectra of the connectivity for two other simple crystals with well-known unit cell without local tetrahedral symmetry, namely face-centered cubic (fcc) and body-centered cubic (bcc) crystals (Fig. 4c) was determined. Despite the fact that the coordination numbers are in these systems are 8 (fcc) and 12 (bcc), the same multiplicity (6) is derived as for cubic diamond.

Finally, the effect of the shape of the unit cell on the Laplace spectra was investigated. The comparison of the Laplace spectra of oxygen connectivity between the cubic and the hexagonal ice can be found in Fig. 4d. For the cubic ice the multiplicity of the two smallest positive eigenvalue is 6 and 12 as we obtained for the other cubic crystal systems (c.f. Fig. 4a and 4c). On the other hand, in the case of not cubic cell the pattern of Laplace spectra at low $\lambda$ values significantly changes. The multiplicity of the first three peaks (eigenvalues) are 3, 3, and 2, respectively. The results suggest that the multiplicity of the first nonzero value of the Laplace spectra depends only on the unit cell parameter.

## 3.3 Effect of the periodic boundary conditions

In the previous section we indirectly showed that the shape (and other related properties) of the Laplace spectra strongly depends on the characteristics of the unit cell. Therefore, it can be presumed that it also depends on the periodic boundary conditions (pbc). To verify that the pbc were eliminated one by one. In this case, we introduced artificially free surface in the system, to do this, we destroyed the orientation equivalence in x, y or z direction. In these cases, the average neighbouring number is decreased about 0.2-0.4.

The result can be seen for cubic diamond and liquid water in Fig. 5. Black sign shows the calculated eigenvalues with periodic boundary conditions. Red, blue and green symbols indicate that the pbc terminated in x direction, in x,y directions, and in x,y,z directions, respectively. As a result, the band gradually disappears. The pattern of the Laplace spectra is decayed step by step. This also proves that the cubic pbc has a detectable effect on the spectra.

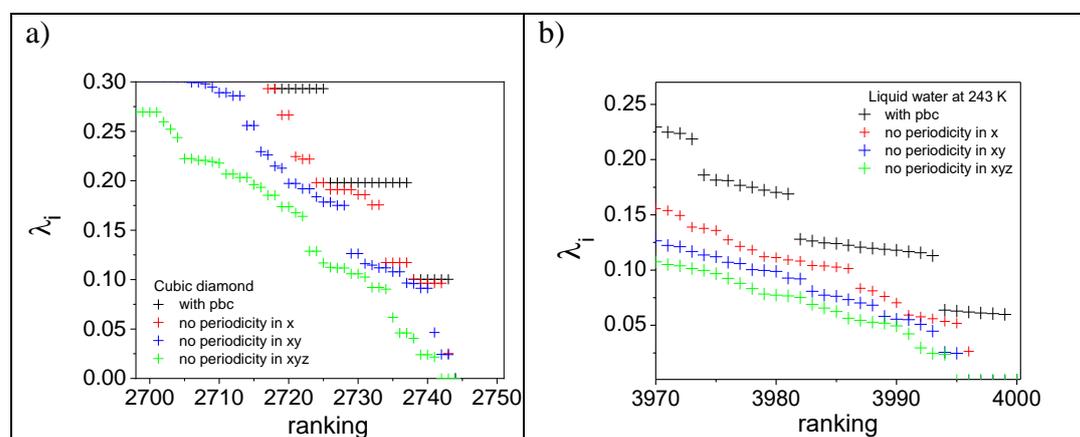

**Figure 5.** The effect of periodic boundary conditions on eigenvalues of Laplace spectra a) for cubic diamond and b) for liquid water at 243 K.

## 3.4 Possible consequences of the six eigenvalues for structural properties of liquid water

In every system, where the unit cell is cubic, the first 6 positive eigenvalues are degenerated, this can be associated with the existence of groups of vertices, which are densely connected between each other while sparsely being connected to the rest of the network. It means that in our systems 6 different ways exist to break defined number of connected bonds leading to the termination of the percolation. In other words, the system is divided into six parts in accordance with the number of eigenvalues. Figure 6 shows representative decomposition patterns for pure water, and cubic diamond. Molecular aggregates were detected by the spectral clustering method in such a way that every molecule is considered as one unit: the water molecules are displayed by their oxygen atom while the cubic diamond by its carbon atom. It is clear that in the cubic diamond cell any type of microheterogeneity cannot be detected. However, in the case of liquid water in every aggregations or clusters the molecules can be classified in two different manners: core or surface molecules. The "core" molecules form H-bond only with molecule inside of the cluster. The "surface" molecules are connected to each other and the core molecules as well. Figure 6c presents only those molecules which can be found on the surface of the clusters in pure water. In this case, the resulted clusters resemble to „tunnels".

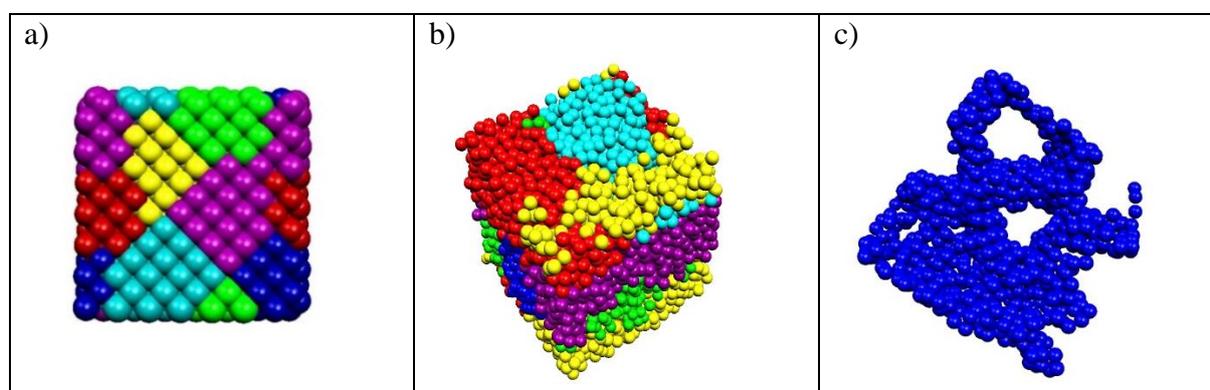

**Figure 6.** A representation of six cluster a) for cubic diamond, b) pure liquid water (243 K) and c) the surface molecules of pure liquid water (243 K).

To connect our results arising from spectral clustering calculations with structural properties of liquid water, the H-bond number distribution was calculated for both the "surface" and the "core" molecules. Results related to the pure water system are shown in Fig. 7. At 298 K the three and the four bonded molecules equally occur inside the cluster ("core" part), while for "surface" molecules the four bonded state is the more favourable. The "surface" type water molecules have a significantly larger tetrahedral nature, than the "core" ones. The average H-bond neigbours number and the fraction of four-bonded (f4) molecules in the case of bulk, surface and core molecules are presented in Table 1.

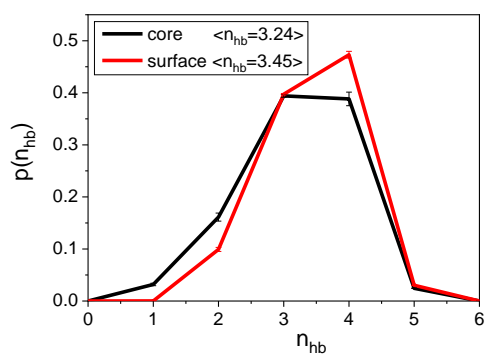

**Figure 7.** Hydrogen bond number distributions for "core" and "surface" molecule in water at different temperature at system size 4000.

|   | $n_{hb}$ | $n_{hb}$ of surface molecules | $n_{hb}$ of core molecules | f4 of surface molecules | f4 of core molecules |
|---|---|---|---|---|---|
| 1 | 3.36 | 3.46 | 3.24 | 0.49 | 0.40 |
| 2 | 3.64 | 3.71 | 3.58 | 0.61 | 0.51 |
| 3 | 3.41 | 3.52 | 3.30 | 0.50 | 0.41 |
| 4 | 3.69 | 3.77 | 3.64 | 0.64 | 0.57 |

**Table 1.** Characteristic values of H-bonded network. (1. $r_{OH}$ <2.5 Å, -12 KJ/mol; 2. $r_{OH}$ <2.5 Å, O…OH < 30.0°; 3. $r_{OO}$ <3.5 Å, -12 KJ/mol; 4. $r_{OO}$ <3.5 Å, O…OH < 30.0°.)

The average number of H-bond is significantly larger (about 0.1-0.23 depending on the H-bond definition c.f. Table S1.) for the molecules on the "surface" than in the "core" part of the clusters. The "surface/core" ratio for pure water system of 4000 molecules at 298 K is about 0.24-0.25, but this ratio slightly depends on the system size. For 8000 molecules this ratio is smaller to some extent, it is approximately 0.19-0.20.

We calculated the tetrahedrality of the two different types of the water molecules for characterizing more deeply the structural difference between the two states. These distributions are presented in Fig. 8a. This figure clearly demonstrates that there is a more pronounced tetrahedral order for the surface than the core type of molecules. The two versions of the calculated local structural index are presented in Fig. 8b and 8c. At low LSI values also show the differences between the "core" and "surface" molecules.

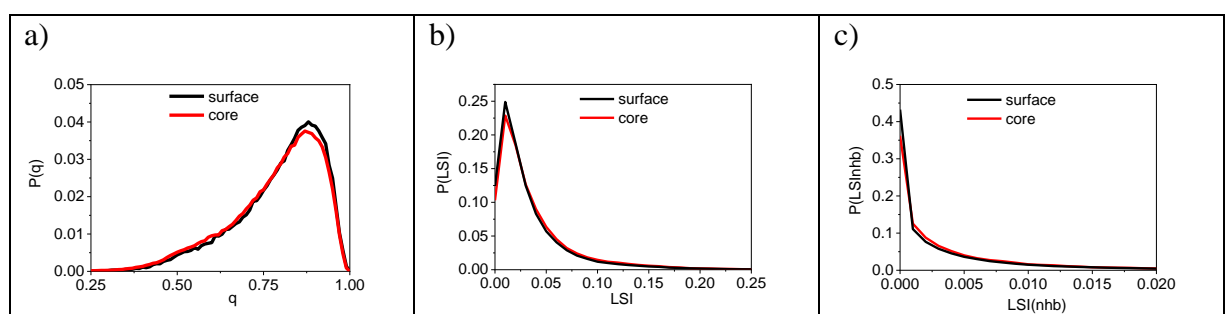

**Figure 8.** a) Probability distribution of tetrahedrality in bulk water at 298 K for surface and core type of molecules. b) Probability distribution of LSI and c) LSI($n_{hb}$) in bulk water at 298 K for surface and core type of molecules.

The value of LSI($n_{hb}$) is significantly smaller than the LSI. This difference can be easily explained by the different in the closest normalized OO distribution functions. The main difference arises from the contribution of 5$^{th}$ neighbor. It is also clearly presented that we take into account in the "classical LSI" calculation the 0.65 fraction of 5$^{th}$ neighbor (Fig. 9).

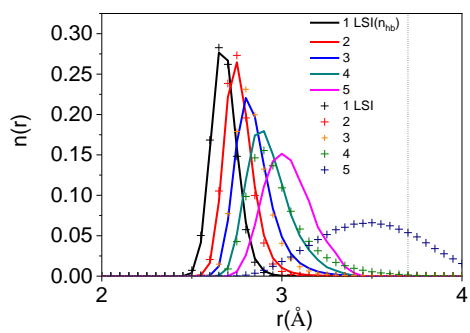

**Figure 9.** The normalized O-O distribution functions of the n$^{th}$ neighbour molecules using two different definitions.

Additionally, we investigated how the calculated $\Delta(r_n) = r(n+1)_{OO} - r(n)_{OO}$ value (for LSI calculation) depends on the $r(n)_{OO}$ value. This two-dimensional distribution is showed in Fig.

10. We can detect two domains. At low r value (r< 2.85 Å), where the r value is smaller than the position of first peak of OO radial distribution, the Δ(rn) is significantly smaller than the right side of the first peak (that is the larger r values).

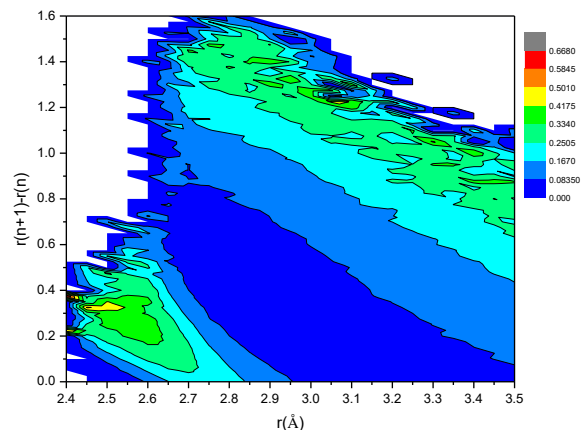

**Figure 10.** Two dimensional distribution of $r(n)_{OO}$ and the corresponding $r(n+1)_{OO}-r(n)_{OO}$.

## 4. Conclusions

In this work we studied the effects of periodic boundary conditions on the community in H-bonded networks based on eigenvectors and eigenvalues analyses of the Laplace spectra applying spectral clustering method. The results suggest that the periodic boundary conditions are always introduce clustering in our systems.

Thus, we are not dealing only with local structural information about the actual structure of the fluid, but an additional distribution, which can be considered as a perturbation arising from periodic boundary conditions. Unfortunately, this effect cannot be eliminated. We hypothesize that these properties of the Laplace spectrum are significantly influenced by the shape of the simulation box. Ultimately, these results suggest that six phases are always formed in our simulation systems in the cubic box due to the periodic boundary conditions. The H-bonded environment on the surface is organized more tetrahedrally than the "core" molecules in the cluster as we showed using several local order parameters such as tetrahedrality and different LSI approaches. Our results revealed that in the case of simulation based investigations of the possible existence of low density liquid (LDL) and high density liquid (HDL) phases in liquid water is always necessary to take into account the effect of periodic boundary conditions.

**Acknowledgements**

The authors are grateful to the National Research, Development and Innovation Office (NRDIO (NKFIH), Hungary) for financial support via grants Nos. K 124885 and FK 128656. Sz. Pothoczki acknowledges that this project was supported by the János Bolyai Research Scholarship of the Hungarian Academy of Sciences.

**References:**

[1] I. J. Farkas, I. Derényi, A.-L. Barabási and T. Vicsek, Spectra of "real-world" graphs: Beyond the semicircle law. Phys. Rev. E 64, 026704 (2001).

[2] Y. Tang, F. Qian, H. Gao, J. Kurths, Synchronization in complex networks and its application: a survey of recent advances and challenge. Annual Reviews in Control 38,184–198 (2014).

[3] A. Barrat, M. Barthélemy, A. Vespignani, Dynamical Processes on Complex Networks. Cambridge University Press (2008).

[4] J. Jost, Dynamical networks. in: F. Jianfeng, J. Jost, Q. Minping (Eds.), Networks: From Biology to Theory. Springer, London, (2007).

[5] M. Newman, Networks: An Introduction. Oxford University Press (2010).

[6] S. Navlakha, C. Kingsford, The power of protein interaction networks for associating genes with diseases. Bioinformatics 26, 1057–1063 (2010).

[7] L. da F. Costa, F. A. Rodrigues, G. Travieso, P. R. Villas Boas, Characterization of complex networks: a survey of measurements. Adv. Phys. 56, 167–242 (2007).

[8] T. Verma, N. A. M. Araujo, H.J. Herrmann, Revealing the structure of the world airline network. Sci. Rep. 4, 5638 (2015).

[9] E. Bullmore, O. Sporns, Complex brain networks: graph theoretical analysis of structural and functional systems. Nat. Rev. Neurosci. 10, 186–198 (2009).

[10] G. A. Pagani, M. Aiello, The power grid as a complex network: a survey. Physica A 392, 2688–2700 (2013).

[11] P. Van Mieghem, Graph Spectra for Complex Networks. Cambridge University Press (2010).

[12] A. Blumen, A. Jurjiu, Multifractal spectra and the relaxation of model polymer networks. J. Chem. Phys. 116, 2636–2641 (2002).

[13] J.-H. Choi, M. Cho, Ion aggregation in high salt solutions. IV. Graph-theoretical analyses of ion aggregate structure and water hydrogen bonding network. J. Chem. Phys. 143, 104110 (2015).

[14] J.-H. Choi, M. Cho, Ion aggregation in high salt solutions. V. Graph entropy analyses of ion aggregate structure and water hydrogen bonding network. J. Chem. Phys. 144, 204126 (2016).

[15] J.-H. Choi, M. Cho, Ion aggregation in high salt solutions. VI. Spectral graph analysis of chaotropic ion aggregates. J. Chem. Phys. 145, 174501 (2016).

[16] J.-H Choi, H. Ran Choi, J. Jeon, M. Cho, Ion aggregation in high salt solutions. VII. The effect of cations on the structures of ion aggregates and water hydrogen-bonding network. J. Chem. Phys. 147, 154107 (2017).

[17] J. A. B. da Silva, F. G. B. Moreira, V. M. L. dos Santos, R. L. Longo, On the hydrogen bond networks in the water–methanol mixtures: topology, percolation and small-world. Phys. Chem. Chem. Phys. 13, 6452-6461 (2011).

[18] A. B. da Silva, F. G. B. Moreira, V. M. L. dos Santos, R. L. Longo, Hydrogen bond networks in water and methanol with varying interaction strengths. Phys. Chem. Chem. Phys. 13, 593-603 (2011).

[19] V. M. L. dos Santos, F. G. B. Moreira, R. L. Longo, Topology of the hydrogen bond networks in liquid water at room and supercritical conditions: a small-world structure. Chem. Phys. Lett. 390, 157-161 (2004).

[20] I. Bakó, I. Pethes, Sz. Pothoczki, L. Pusztai, Temperature dependent network stability in simple alcohols and pure water: The evolution of Laplace spectra. J. Mol. Liq. 273, 670–675 (2019.)

[21] I. Bakó, A. Bencsura, K. Hermannson, S. Bálint, T. Grósz, V. Chihaia, J. Oláh, Hydrogen bond network topology in liquid water and methanol: a graph theory approach. Phys. Chem. Chem. Phys. 15, 15163–15171(2013).

[22] J. L. F. Abascal and C. Vega "A general purpose model for the condensed phases of water: TIP4P/2005", Journal of Chemical Physics, 123 234505 (2005)

[23] U. Von Luxburg, A tutorial on spectral clustering. Stat. Comput. 17, 395-416 (2007).

[24] U. Von Luxburg, M. Belkin, O. Bousquet, Consistency of spectral clustering. The Annals of Statistics 36, 555-586 (2008).

[25] A. J Seary, W. D. Richards, Partitioning networks by eigenvectors. Proceedings of the International Conference on Social Networks 1, 47-58 (1995).

[26] D. Cvetkovic, S. Simic, Graph spectra in Computer Science. Linear Algebra Appl. 434, 1545–1562 (2011).

[27] S. White, P. Smyth, A spectral clustering approach to finding communities in graphs. Proceedings of the 2005 SIAM International Conference on Data Mining 274-285 (2005).

[28] S. E. Schaeffer, Graph clustering. Computer Science Review 1, 27–64 (2007).

[29] A. Y. Ng, M. I. Jordan, Y. Weiss, On spectral clustering: Analysis and an algorithm. Proceedings of the 14th International Conference on Neural Information Processing Systems: Natural and Synthetic 849-856, (2001).

[30] M. E. J. Newman, Finding community structure in networks using the eigenvectors of matrices. Phys. Rev. E 74, 036104 (2006).

[31] Graph Partitioning. Edited by Charles-Edmond Bichot and Patrick Siarry. Wiley, New York (2011).

[32] F. R. K. Chung, Spectral Graph Theory. Chapter 2.2. A.M.A. CBMS, Providence, Rhode Island (1997).

[33] O. Mishima, H. E. Stanley, Decompression-Induced Melting of Ice IV and the Liquid-Liquid Transition in Water. Nature 392,164−168 (1998).

[34] O. Mishima, Liquid-Liquid Critical Point in Heavy Water. Phys. Rev. Lett. 85, 334−336 (2000).

[35] I. Zhovtobriukh, N. A. Besley, T. Fransson, A. Nilsson, Lars G. M. Pettersson, Relationship between x-ray emission and absorption spectroscopy and the local H-bond environment in water. J. Chem. Phys. 148, 144507 (2018).

[36] G. Camisasca, D. Schlesinger, I. Zhovtobriukh, G. Pitsevich, Lars G. M Pettersson, A proposal for the structure of high- and low-density fluctuations in liquid water. J. Chem. Phys. 151, 034508 (2019).

[37] P. Wernet, D. Nordlund, U. Bergmann, M. Cavalleri, M. Odelius, H. Ogasawara, L. Å. Näslund, T. K. Hirsch, L. Ojamäe, P. Glatzel et al., The structure of the first coordination shell in liquid water, Science 304, 995–999 (2004).

[38] A. Nilsson, D. Nordlund, I. Waluyo, N. Huang, H. Ogasawara, S. H. Kaya, U. Bergmann, L.- Å. Näslund, H. Öström, P. Wernet, K. Andersson, T. Schiros, L. G. M. Pettersson, X-ray Absorption Spectroscopy and X-ray Raman Scattering of Water: An Experimental View. J. Electron Spectrosc. Relat. Phenom 177, 99−129 (2010).

[39] J. R. Scherer, M. K. Go, S. Kint, Raman spectra and structure of water from −10 to 90.deg. J. Phys. Chem. 78, 1304–1313 (1974).

[40] J. A. Sellberg, T. A. McQueen, H. Laksmono, S. Schreck, M. Beye, D. P. DePonte, B. Kennedy, D. Nordlund, R. G. Sierra, D. Schlesinger, T. Tokushima, I. Zhovtobriukh, S. Eckert, V. H. Segtnan, H. Ogasawara, K. Kubicek, S. Techert, U. Bergmann, G. L. Dakovski, W. F. Schlotter, Y. Harada, M. J. Bogan, P. Wernet, A. Föhlisch, L. G. M. Pettersson, A. Nilsson, X-ray Emission Spectroscopy of Bulk Liquid Water in "No-man's Land". J. Chem. Phys. 142, 044505 (2015).

[41] J. C. Palmer, R. Car, P. G. Debenedetti, The liquid-liquid transition in supercooled ST2 water: a comparison between umbrella sampling and well-tempered dynamics. Faraday Discuss. 167, 77-94 (2013).

[42] A. Taschin, P. Bartolini, R. Eramo, R. Righini, R. Torre, Evidence of Two Distinct Local Structures of Water from Ambient to Supercooled Conditions. Nat. Commun. 4, 2401 (2013).

[43] M. Matsumoto, T. Yagasaki, H. Tanaka, A Bayesian approach for identification of ice Ih, ice Ic, high density, and low density liquid water with a torsional order parameter. J. Chem. Phys. 150, 214504 (2019).

[44] F. Martelli, Unravelling the contribution of local structures to the anomalies of water: The synergistic action of several factors. J. Chem. Phys. 150, 094506 (2019).

[45] N. Ansari, R. Dandekar, S. Caravati, G.C. Sosso, A. Hassanali, High and low density patches in simulated liquid water. J. Chem. Phys. 149, 204507 (2018).

[46] D. T. Limmer, D. Chandler, The putative liquid-liquid transition is a liquid-solid transition in atomistic models of water. J. Chem. Phys. 135, 134503 (2011).

[47] D. T. Limmer and D. Chandler, The putative liquid-liquid transition is a liquid-solid transition in atomistic models of water. II. J. Chem. Phys. 138, 214504 (2013).

[48] J. D. Smith, C. D. Cappa, K. R. Wilson, R. C. Cohen, P. L. Geissler, R. J. Saykally, Unified description of temperature-dependent hydrogen-bond rearrangements in liquid water. PNAS 102, 14171-14174 (2005).

[49] K. Binder, Simulations Clarify When Supercooled Water Freezes into Glassy Structures. PNAS 111, 9374−9375 (2014).

[50] J. Niskanen, M. Fondella, C. J. Sahle, S. Eckert, R. M. Jay, K. Gilmore, A. Pietzsch, M. Dantz, X. Lu, D. E. McNally, T. Schmitt, V. Vaz da Cruz, V. Kimberg, F. Gelmukhanov, A. Föhlisch Compatibility of quantitative X-ray spectroscopy with continuous distribution models of water at ambient conditions. PNAS 116, 4058-4063 (2019).

[51] G. N. I. Clark, G. L. Hura, J. Teixeira, Alan K. Soper, T. Head-Gordon Small-angle scattering and the structure of ambient liquid water. PNAS 107, 14003–14007 (2010).

[52] S. Sastry, P. G. Debenedetti, F. Sciortino, H. E. Stanley, Singularity-free interpretation of the thermodynamics of supercooled water. Phys. Rev. E 53, 6144-6154 (1996).

[53] A. Banerjee, J. Jost, On the spectrum of the normalized graph Laplacian. Linear Algebra Appl. 428, 3015-3022 (2008).

[54] A. Banerjee, J. Jost, Spectral characterization of network structures and dynamics. Dynamics on and of Complex Networks, 117-132 (2009).

[55] P. N. McGraw, M. Menzinger, Laplacian spectra as a diagnostic tool for network structure and dynamics. Phys. Rev. E, 77, 031102 (2008).

[56] S. Hata, H. Nakao, Localization of Laplacian eigenvectors on random networks. Sci. Rep. 7, 1121 (2017).

[57] A. Julaiti, B. Wu, Z. Zhang, Eigenvalues of normalized Laplacian matrices of fractal trees and dendrimers: analytical results and applications. J. Chem. Phys. 138, 204116 (2013).

[58] M. Fiedler, A property of eigenvectors of nonnegative symmetric matrices and its applications to graph theory. Czech. Math. J. 25, 619-633 (1975).

[59] J. Friedman, Some geometric aspects of graphs and their eigenfunctions. Duke Math J. 69, 487-525 (1993).

[60] N. M. M. de Abreu, Old and new results on algebraic connectivity of graphs. Linear Algebra Appl. 423, 53–73 (2007).

[61] D. van der Spoel, E. Lindahl, B. Hess, G. Groenhof, A. E. Mark, H. J. C. Berendsen, GROMACS: Fast, flexible, and free, J. Comput. Chem. 26, 1701 (2005).

[62] P.-L. Chau, A. J. Hardwick, A New Order Parameter for Tetrahedral Configurations. Mol. Phys. 93, 511−518 (1998).

[63] J. R. Errington, P. G. Debenedetti, Relationship Between Structural Order and the Anomalies of Liquid Water. Nature 409, 318-21 (2001).